\documentclass[
    ,final            
  ]
  {aipproc}

\layoutstyle{8x11double}

\def\P{\mathcal{P}}
\def\tr{\mbox{tr}}

\def\fig{Fig.}

\begin{document}

\title{Vortex content of calorons\\ and deconfinement mechanism
\footnote{talk given by FB at `Quark Confinement And The Hadron Spectrum IX', Madrid, Sept. 2010.}
}


\author{Falk Bruckmann}{
  address={Institut f\"ur Theoretische Physik, Universit\"at
  Regensburg, D-93040 Regensburg, Germany}
}

\author{Ernst-Michael\ Ilgenfritz}{
  address={Institut f\"ur Physik, Humboldt-Universit\"at, D-12489 Berlin, Germany}
}

\author{Boris Martemyanov}{
  address={Institute for Theoretical and Experimental Physics, B.\ Cheremushkinskaya 25, 117259 Moscow, Russia}
}

\author{Bo Zhang}{
  address={Institut f\"ur Theoretische Physik, Universit\"at
  Regensburg, D-93040 Regensburg, Germany}
}


\begin{abstract}
We reveal the center vortex content of SU(2) calorons and ensembles of them. While one part of the vortex connects the constituent dyons of a single caloron, another part is predominantly spatial and can be related to the twist that exists in the caloron gauge field. The latter part depends strongly on the caloron holonomy and degenerates to a plane between the dyons when the asymptotic Polyakov loop is traceless. Correspondingly, the spatial vortex in caloron ensembles is percolating in this case. This finding fits perfectly in the confinement scenario of vortices and shows that calorons are suitable to facilitate the vortex (de)confinement mechanism.
\end{abstract}

\maketitle

Calorons, instantons at finite temperature, possess dyons (electrically charged magnetic
monopoles, $N$ for gauge group SU($N$)) as constituents and hint at confinement in various ways, see  \cite{Bruckmann:2007d_Diakonov:2009} and references therein. 

\begin{figure}[!b]
\begin{minipage}{0.9\linewidth}
 \includegraphics[width=0.8\linewidth]{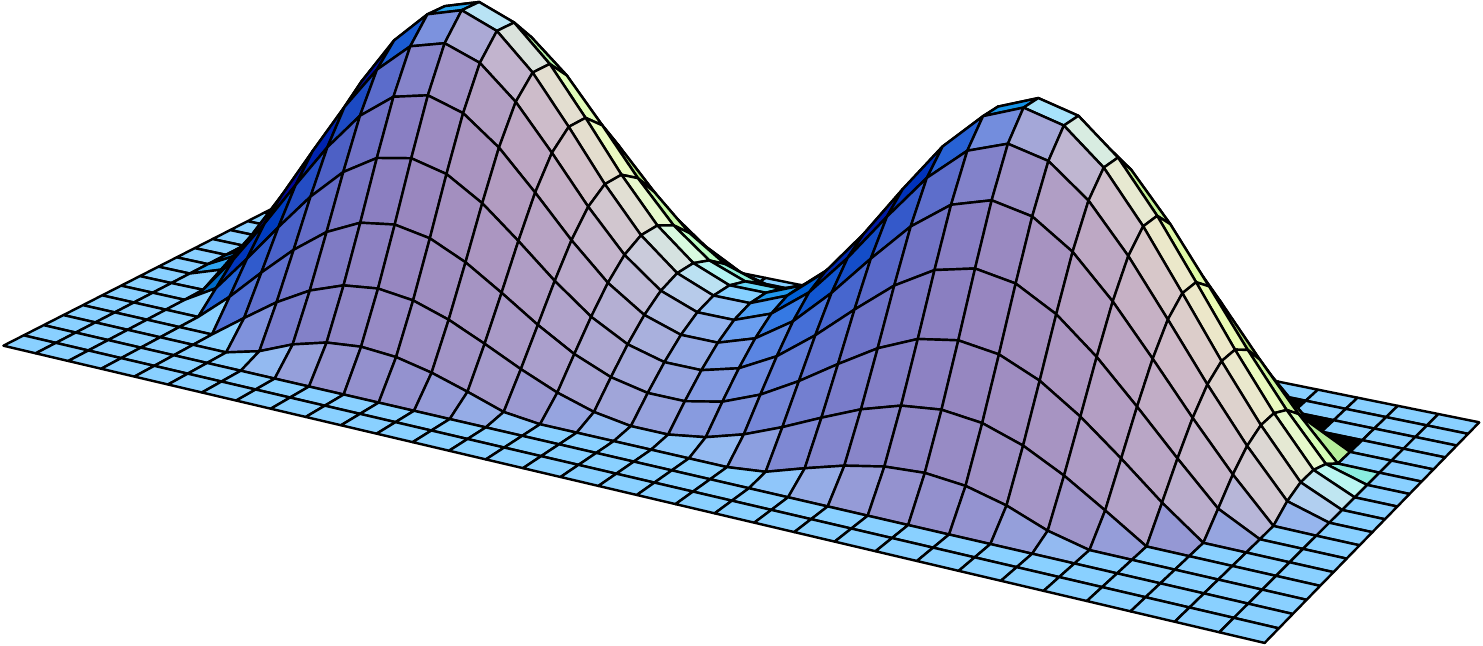}\\

\vspace{-0.5cm}
 \includegraphics[width=0.8\linewidth]{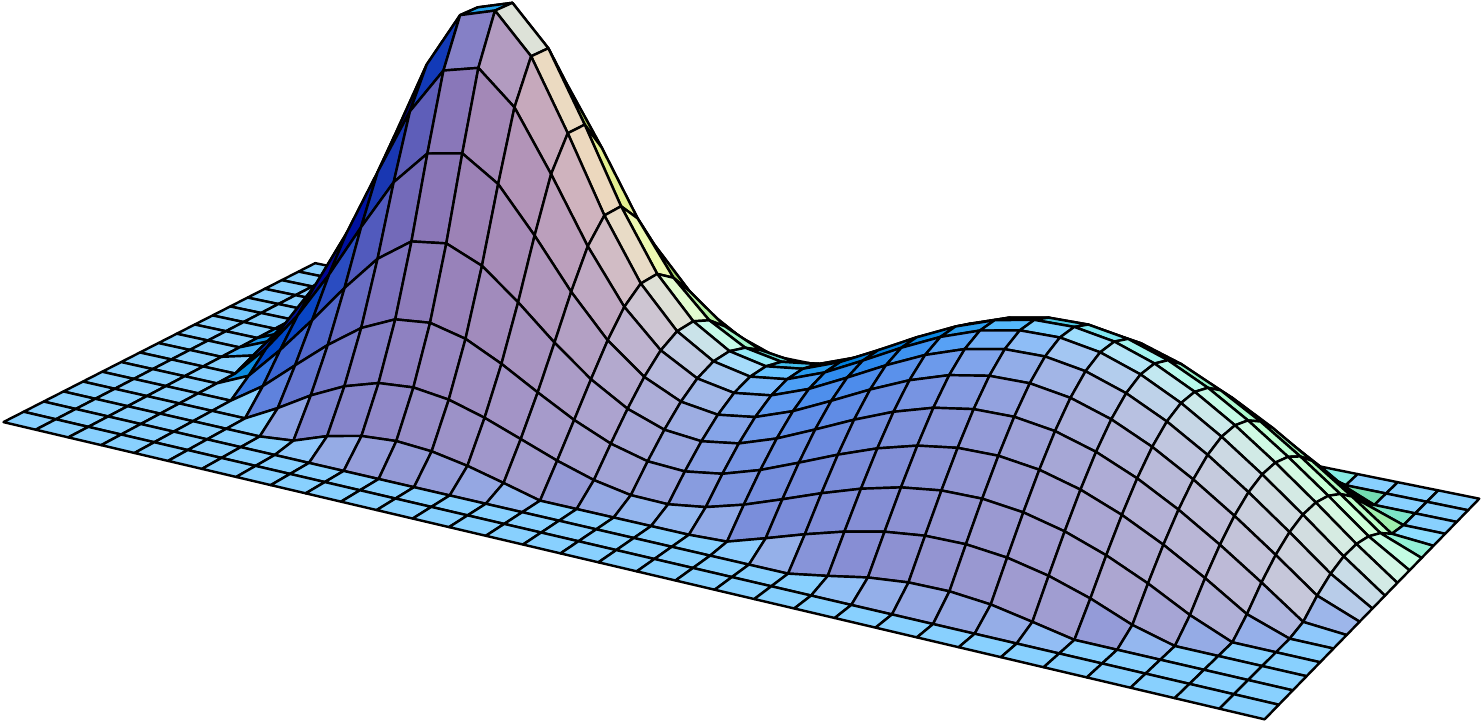}
\end{minipage}
\caption{Action/topological density of $SU(2)$ calorons with their dyon constituents for traceless holonomy $\tr\,\P_\infty=0$  (top) and close to trivial holonomy $\tr\,\P_\infty/2=1/\sqrt{2}$ (bottom). 
}
\source{\cite{Kraan:1998}}
\label{fig calorons}
\end{figure}

Center vortices, on the other hand,  have been proposed since a long time as QCD excitations relevant for confinement and other nonperturbative effects, see  \cite{Greensite:2003} and references therein. Wilson loops linked to a vortex take values of nontrivial center elements. Hence percolating vortices disorder Wilson loops giving rise to the area law. Vortices are 2D objects, but are obtained in QCD only through a center projection.


\begin{figure}[!b]
\begin{minipage}{0.9\linewidth}

\vspace{-0.2cm}
 \qquad\qquad\qquad\qquad\quad
\includegraphics[width=0.55\linewidth]{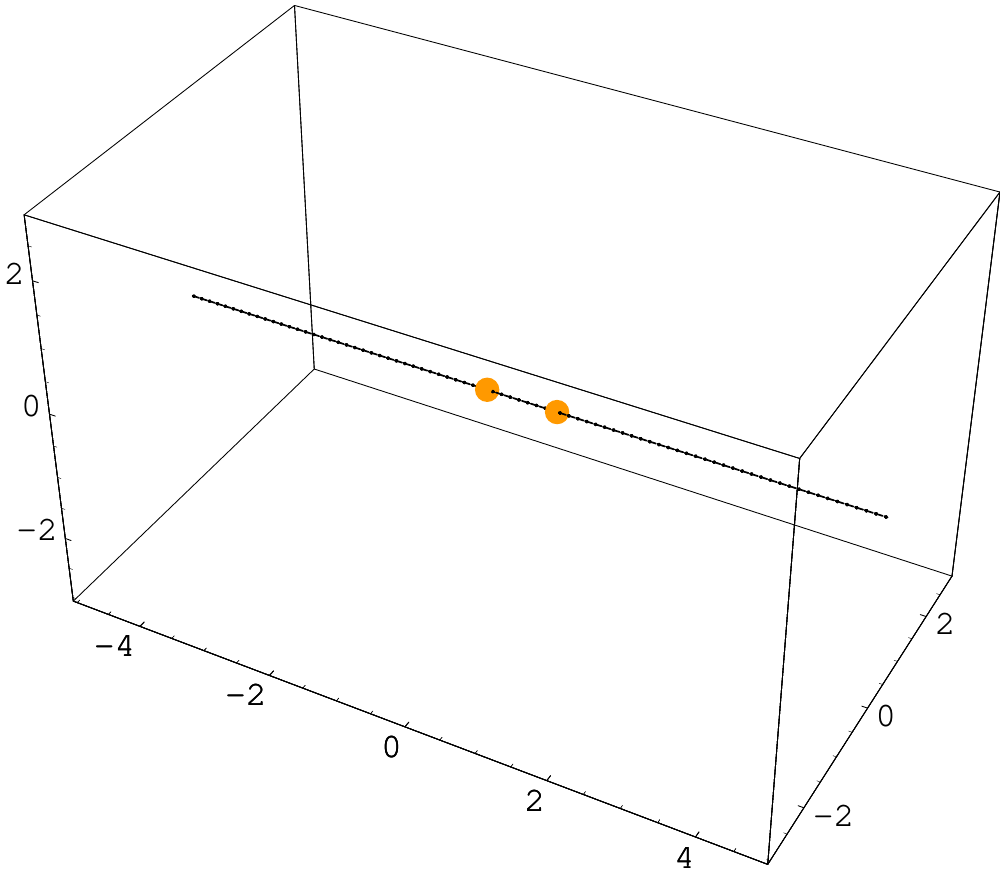}

\vspace{-1.4cm}
\hspace{-0.5cm}
\includegraphics[width=0.55\linewidth]{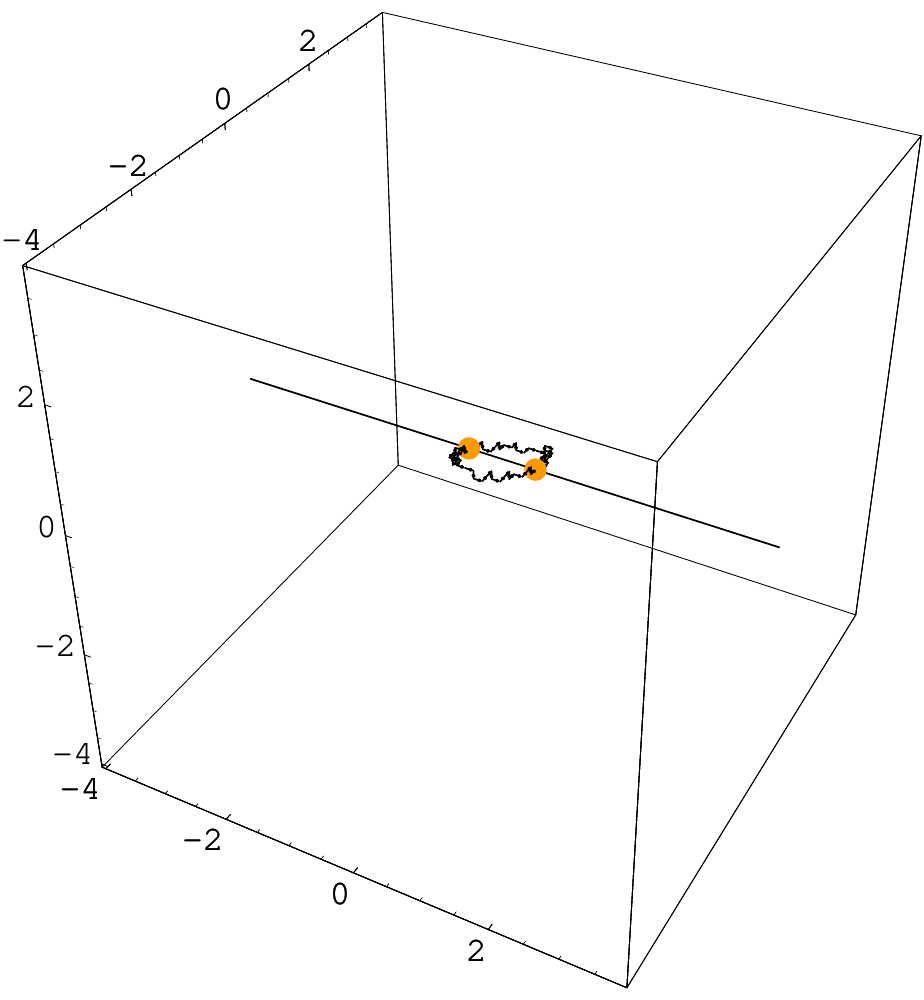}
\end{minipage}
\caption{The space-time vortex, line-like at fixed time, is spanned between dyons (marked by dots). Top: in an elongated box the vortex coincides with the axis connecting the dyons. Bottom: in a cubic box two vortex lines connect the dyons in the caloron's interior (with the axis to guide the eye). In both cases the vortex extends in time and the flux along the vortex flips at the dyons.
}
\label{fig st vortex}
\end{figure}

We will study the connection between vortices and SU(2) calorons using Laplacian center
gauge \cite{Forcrand:2001} on these classical backgrounds (semi-analytically, discretized on a lattice). For all details we refer to  \cite{Bruckmann:2009c}.

The two dyon constituents, as shown in \fig~\ref{fig calorons}, depend on the holonomy $\P_\infty$, their asymptotic Polyakov loop. Identifying the trace of the latter with the ave\-rage traced Polyakov loop as the order
parameter of (de)confinement, the properties of the dyons are sensitive to the phase: in
the confined phase, the dyons are of same mass (fraction of topological charge) whereas
in the high temperature phase calorons consist of one heavy plus one light dyon. We use the holonomy 
to impose the actual 
phase 
to the caloron solutions.
Let us first describe the vortex content of a single caloron.

One part of the single calorons' vortex extends in space and time. It contains the constituent dyons as it should be: Laplacian Abelian Gauge \cite{Sijs:1997} detects the dyons as monopole worldlines  \cite{Bruckmann:2009c} 
and is part of our center projection. This so-called `dyon-induced' vortex is depicted in \fig~\ref{fig st vortex}.

We found an ambiguity, depending on the aspect ratio of the box (lattice), in how the obtained vortex surface connects the dyon worldlines. In both cases there are two fluxes towards respectively away from each dyon rendering the vortex surface non-orientable. 

\begin{figure}[!b]
\begin{minipage}{0.9\linewidth}

\vspace{-0.2cm}
\includegraphics[width=0.7\linewidth]{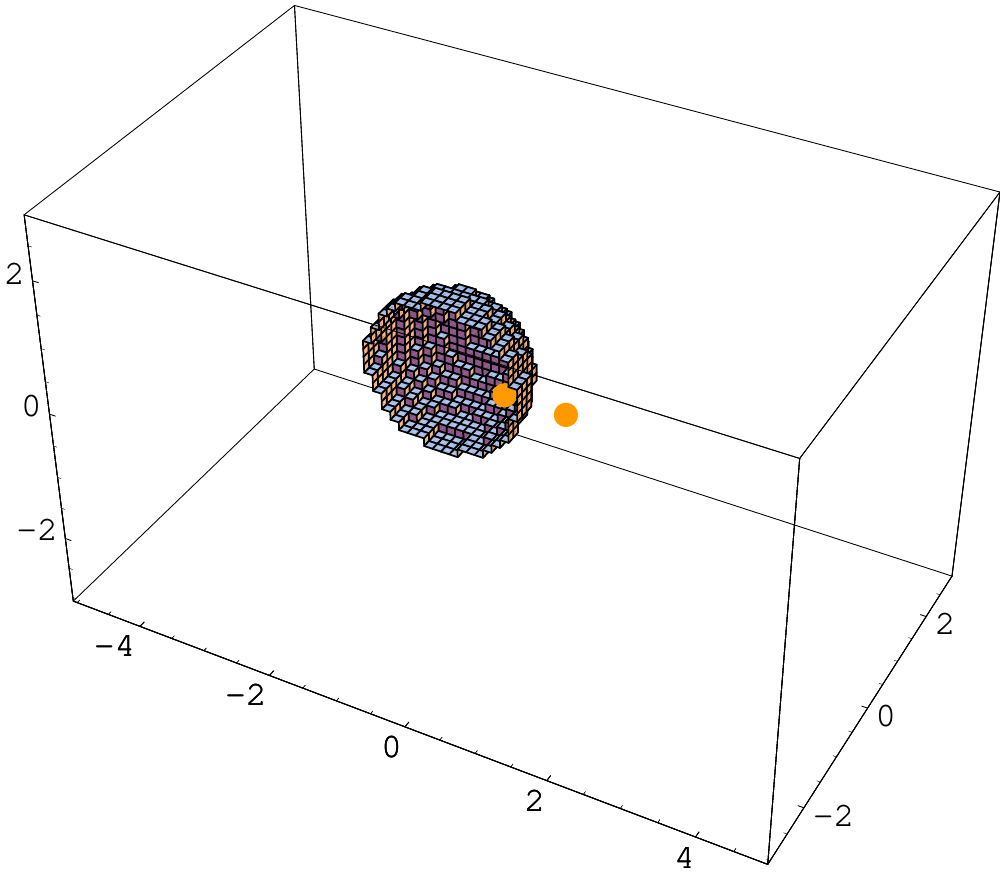}\\

\vspace{-0.7cm}
\includegraphics[width=0.7\linewidth]{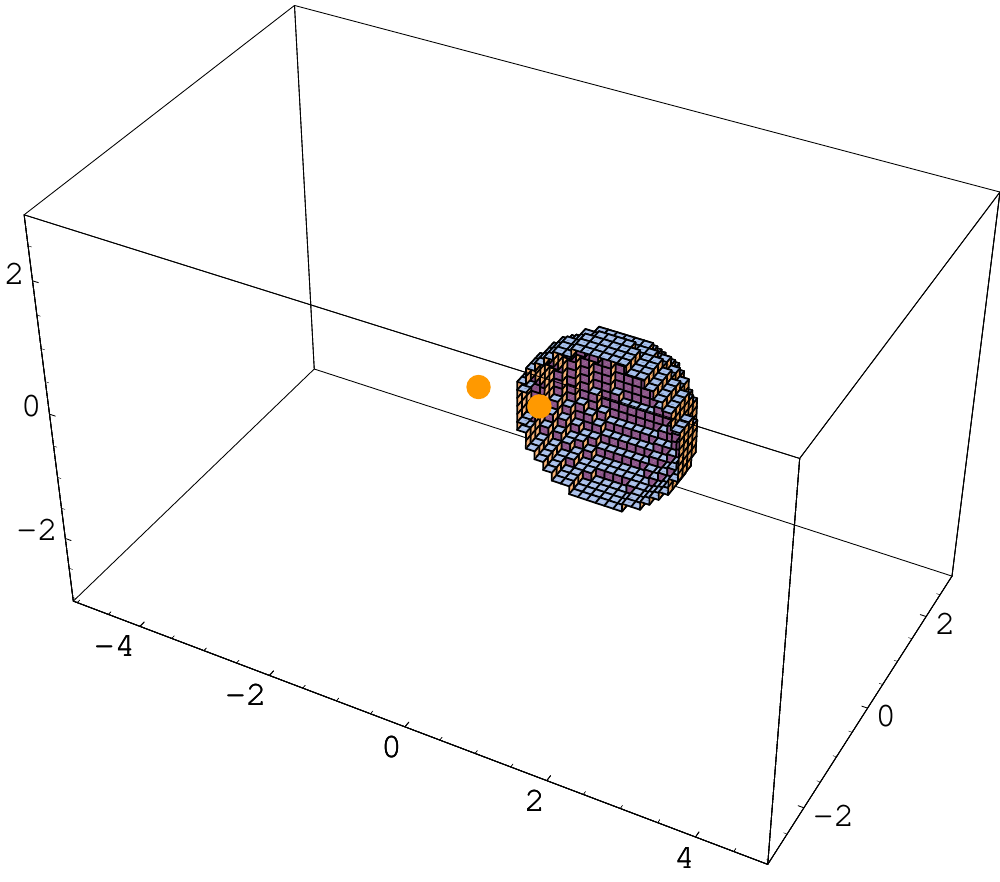}
\end{minipage}
\caption{Purely spatial vortex, a `bubble' (cut) around one dyon (dyons marked by dots) for holonomies $\tr\,\P_\infty/2=\pm 0.43$ (top/bottom) and caloron size parameter $\rho=0.6~\beta$.} 
\label{fig s vortex}
\end{figure}

A second part of the vortex is purely spatial, i.e.\ appears at a fixed time slice. It enclosed one of the dyons by a `bubble', which one depends on the sign of the holonomy (relative to the one of the local Polyakov loop at the dyon), see \fig~\ref{fig s vortex}. 

This vortex can be shown to be induced by the `twist', the  fact that one dyon is gauge rotated with time w.r.t.\ the other dyon in the caloron. The lowest Laplacian modes behave in a similar way around the dyons, from which a winding number argument predicts the existence of a spatial vortex at the boundary between the static and twisting region \cite{Bruckmann:2009c}. 

\begin{figure}[!t]
\includegraphics[width=\linewidth]{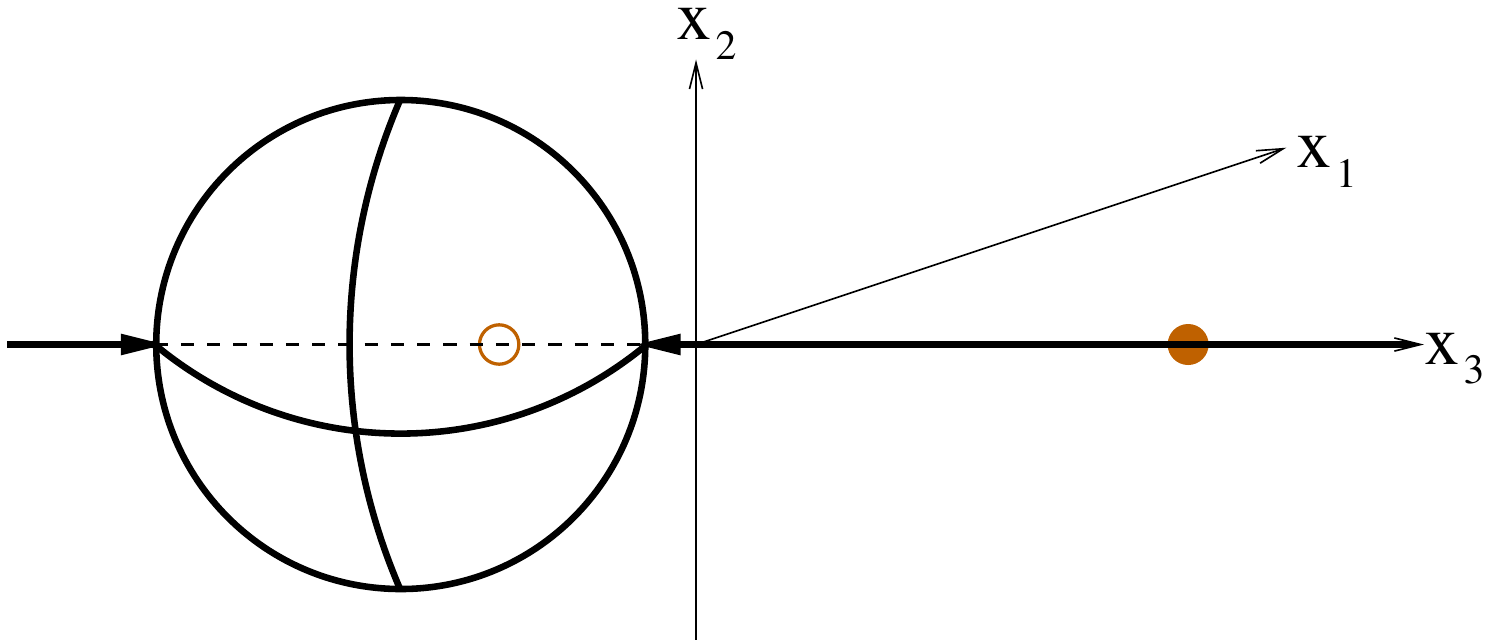}
\caption{Sketch of the intersection of the two vortex parts giving topological charge (present in just one time slice).}
\label{fig intersection}
\end{figure}

Both vortex parts together result in two intersection points as shown schematically in \fig~\ref{fig intersection}. Since at these points the vortex' target space spans all four directions, topological charge is generated. It is known to be a unit $Q=1/2$ at each intersection point \cite{Engelhardt:2000_Reinhardt:2002_Engelhardt:2000a_Bruckmann:2003b}, here with the same sign due to the direction of the fluxes (and non-orientability). Hence the vortex reproduces the topological charge 1 of the background caloron (without writhe).

\begin{figure}[!b]
\begin{minipage}{0.9\linewidth}

\vspace{-0.1cm}
\includegraphics[width=0.7\linewidth]{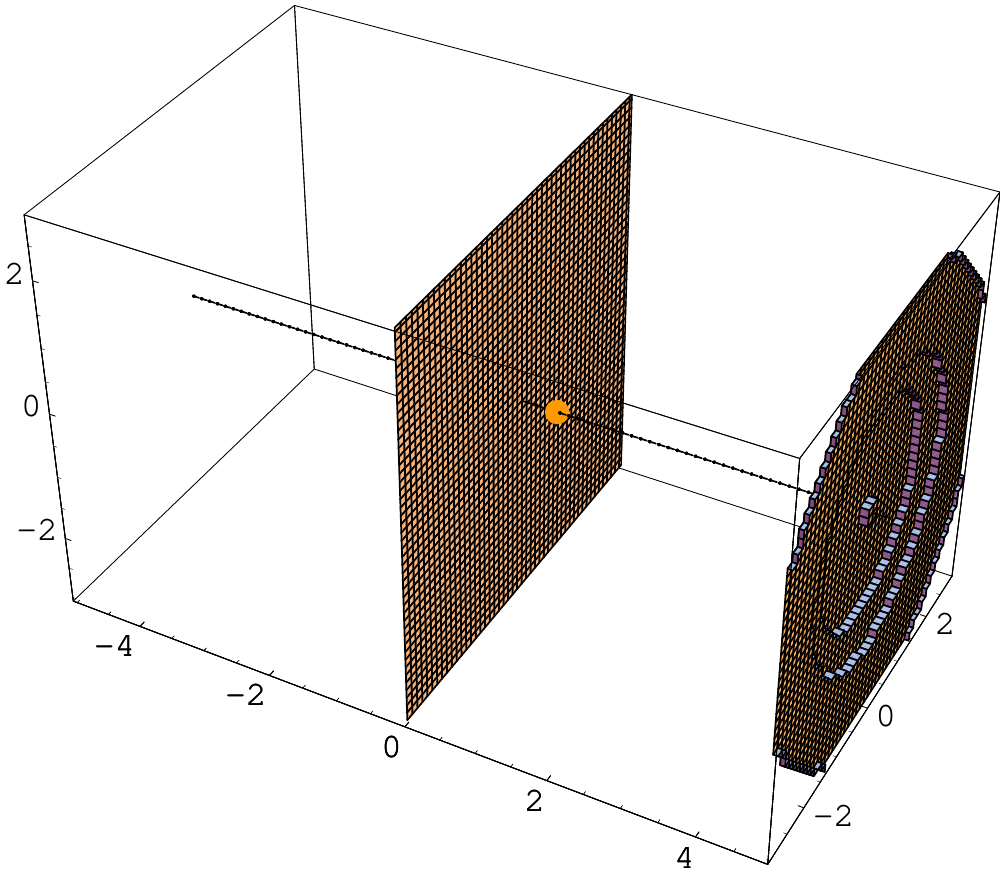}\\

\end{minipage}
\caption{Purely spatial vortex of individual calorons again (cf.\ \fig~\protect\ref{fig s vortex}), here for maximal nontrivial holonomy $\tr\,\P_\infty/2=0$. (The plane near the boundary is an artifact caused by periodic boundary conditions.)} 
\label{fig s vortex again}
\end{figure}

As one of our goals are Polyakov loop correlators, we come back to the spatial vortices induced by twist and the influence of the holonomy. \fig~\ref{fig s vortex again} shows that in the limiting case of traceless holonomy, $\tr\P_\infty=0$ representing the confined phase, the `bubble' degenerates to the midplane between the dyons (also for symmetry reasons). For the other limiting cases $\tr\P_\infty/2\to \pm1$ the `bubble' becomes very small \cite{Bruckmann:2009c} [not shown].

Random superpositions of calorons constitute a model for the QCD vacuum at finite temperature in the semiclassical approach. We use configurations obtained by (non-interacting) superposition of calorons from \cite{Gerhold:2007} with a rather low density of 6 calorons and 6 anti-calorons in a $8\times 64^3$ lattice (with average size $\bar{\rho}=0.6\beta$). We compare configurations  with the same dyon locations only differing in the holonomy in order to illustrate the essential difference between the low and high temperature phase.

\begin{figure}[!t]
\begin{minipage}{0.9\linewidth}
\includegraphics[width=0.67\linewidth]{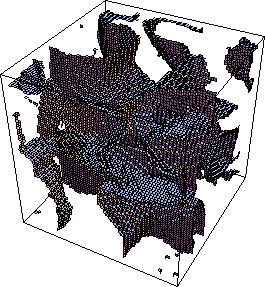}
\includegraphics[width=0.67\linewidth]{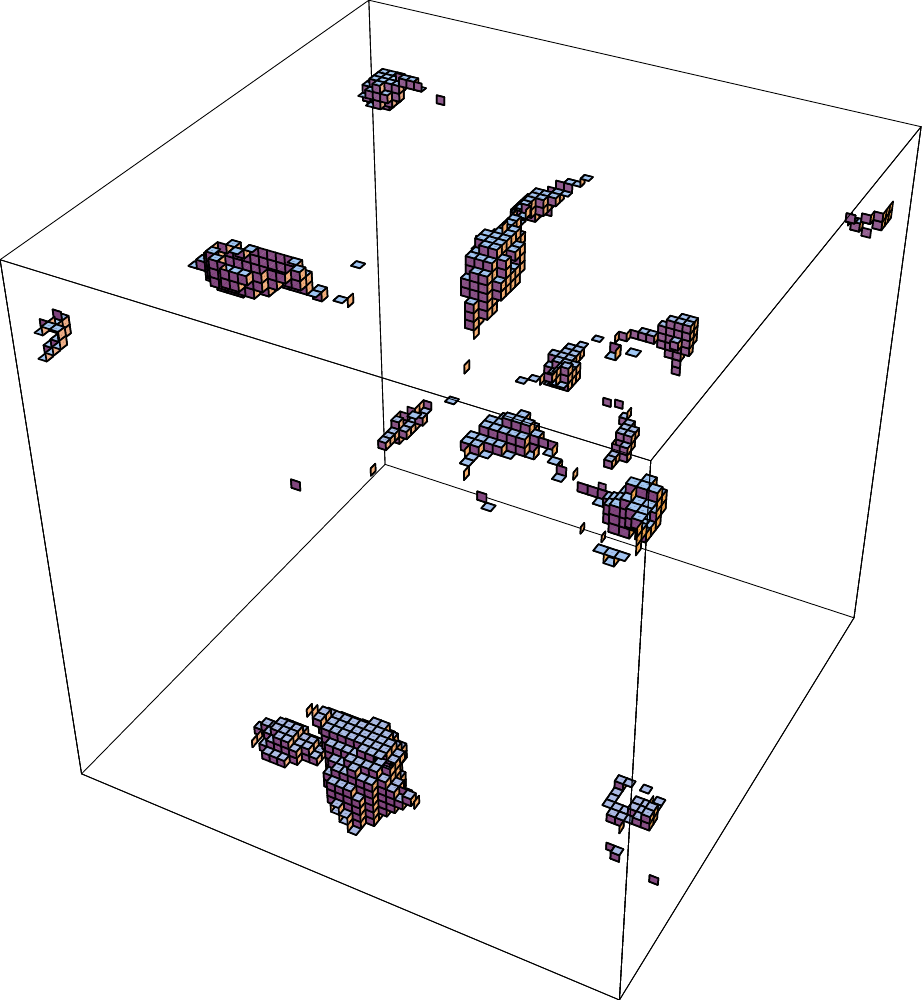}
\end{minipage}
\caption{Purely spatial vortex content of configuration from a caloron gas with maximal nontrivial holonomy $\tr\,\P_\infty/2=0$ (top) representing the low temperature phase and holonomy close to trivial, $\tr\,\P_\infty/2=0.92$ representing the high temperature phase (bottom).} 
\label{fig s vortex gas}
\end{figure}

\begin{figure}[!t]
\begin{minipage}{0.9\linewidth}
\includegraphics[width=0.7\linewidth]{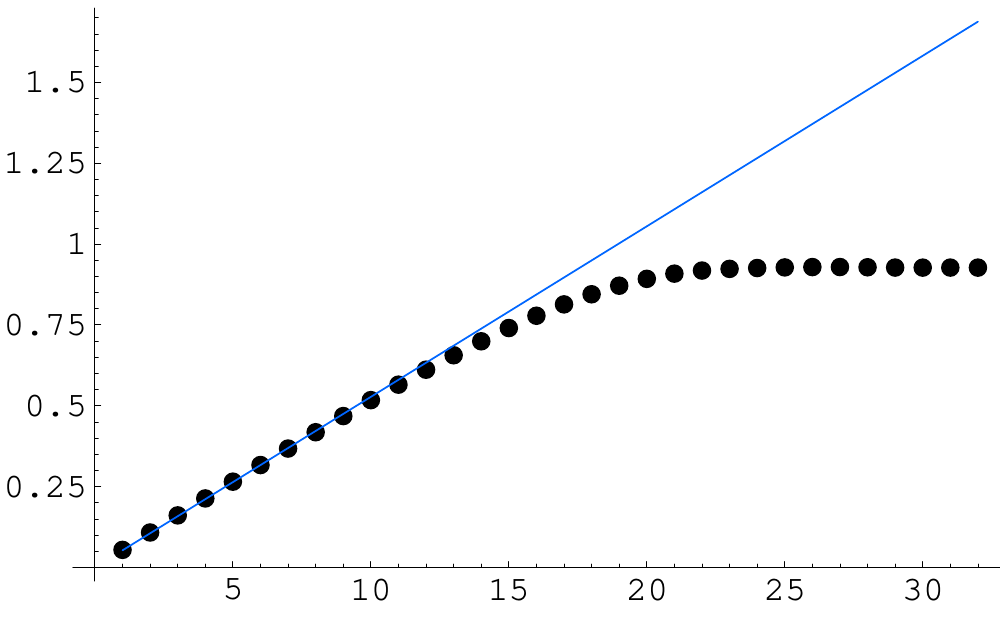}
\includegraphics[width=0.7\linewidth]{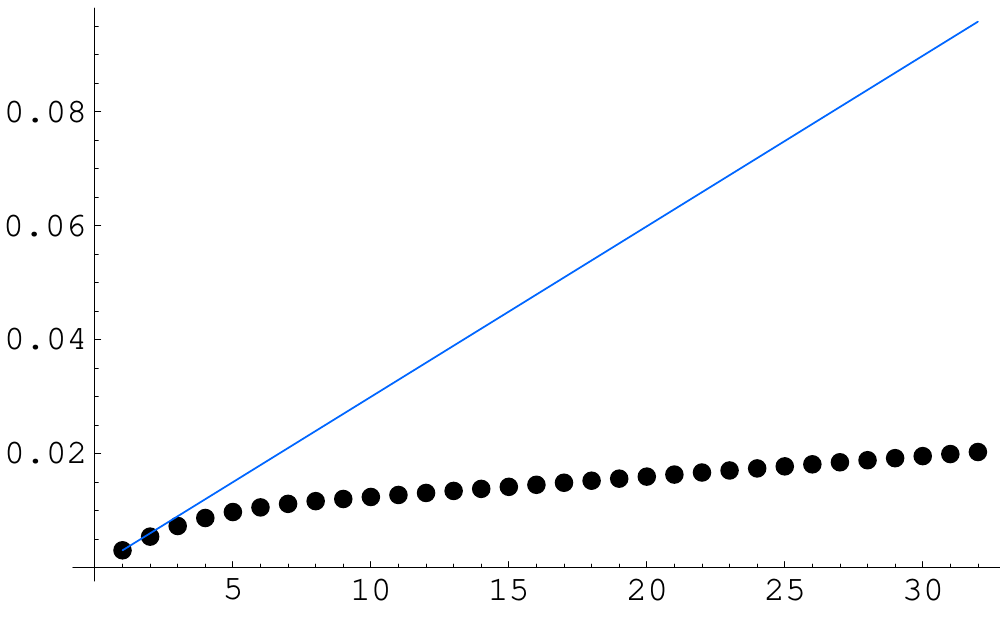}
\end{minipage}
\caption{Interquark potential from Polyakov loop correlators (after projection onto
the center) for the holonomies of \fig~\protect\ref{fig s vortex gas} as a function of distance in `lattice spacings' $a=\beta/8$. Note the different scales.} 
\label{fig string tensions}
\end{figure}

The vortex content of this caloron gas can be described in a first approximation by the superpositions of individual vortices. As a consequence, in the low temperature phase with $\tr \P_\infty= 0$ the midplanes described above touch each other and merge to a large vortex cluster, see \fig~\ref{fig s vortex gas} top panel. Such a percolation of spatial vortices is  exactly the mechanism to yield an area law for space-time Wilson loops \cite{Engelhardt:2000b,Engelhardt:2000c,Greensite:2003}. We demonstrate this by the logarithm of Polyakov loop correlators (`interquark potential') where this disorder generates a linear dependence on the distance, see \fig~\ref{fig string tensions}, top panel. The corresponding string tension can be estimated to $0.4/\beta^2$, which is of same order of magnitude as the string tension from full gauge fields of calorons obtained in \cite{Gerhold:2007}, $200~\mbox{MeV/fm}$, which can be rewritten as $1.24/\beta^2$ at critical temperature.

In contrast, in the high temperature phase with $\tr \P_\infty/2\to 1$ the bubbles stay isolated and do not percolate in three-dimensional space, see \fig~\ref{fig s vortex gas}, bottom panel. The corresponding string tension is suppressed by orders of magnitude, 
see \fig~\ref{fig string tensions}, bottom panel. 


At the same time, the change in the space-time vortices of this caloron gas is much less drastic \cite{Bruckmann:2009c} [not shown]. This means that the spatial Wilson loops behave rather smoothly across the phase transition, as is known to be the case in (quenched) QCD.
These findings are summarized by the tendency of the vortices in the high temperature phase to align in the time-like direction, cf.\ \cite{Engelhardt:2000b}.

We thank Jeff Greensite for discussions sharing his insights into vortices. F.B. and B.Z. have been supported by DFG (2872/4-1).

\end{document}